\documentstyle[preprint,eqsecnum,aps,prb]{revtex}

\begin{document}
\title{Giant anisotropic magnetostriction in Pr$_{0.5}$Sr$_{0.5}$MnO$_{3}$}
\author{R. Mahendiran*, C. Marquina, and M. R. Ibarra}
\address{Departamento de F\'{i}sica y de la Materia Condensada-ICMA, Universidad de\\
Zaragoza-CSIC, Zaragoza-50009, Spain.}
\author{A. Arulraj and C. N. R. Rao}
\address{Chemistry and Physics of Materials Unit, Jawaharlal Nehru Centre for\\
Advanced Scientific Research, Jakkur P. O. Bangalore-560 064, India.}
\author{A. Maignan and B. Raveau}
\address{Laboratoire CRISMAT, ISMRA, Universite de Caen, 6. Boluevard du\\
Mar\'{e}chal Juin, Caen Cedex-14050, France.}
\maketitle

\begin{abstract}
Magnetic, linear thermal expansion (LTE), anisotropic ($\lambda _t$) and
volume ($\omega $) magnetostriction properties of \ Pr$_{0.5}$Sr$_{0.5}$MnO$%
_3$ were investigated. The LTE decreases smoothly from 300 K without a clear
anomaly either around the Curie (T$_C$ = 270 K) or the Neel temperature (T$_N
$ = 100 K) and it exhibits hysteresis over a wide temperature range (60
K-270 K) upon warming. Isothermal magnetization study suggests that 13 \% of
the ferromagnetic phase coexists with 87 \% of the antiferromagnetic phase
at 25 K. The parallel and perpendicular magnetostrictions undergo rapid
changes during the metamagnetic transition. Contrary to the isotropic giant
volume magnetostriction reported in manganites so far, this compound
exhibits a giant anisotropic magnetostriction ($\lambda _t$ $\approx $ 10$%
^{-3}$) and smaller volume ($\omega $ $\approx $ 10$^{-4}$)
magnetostrictions below T$_N$. We suggest that the field induced
antiferromagnetic to ferromagnetic transition is accompanied by a structural
transition from the d$_{x^2-y^2}$ orbital ordered antiferromagnetic
(orthorhombic) to the orbital disordered ferromagnetic (tetragonal) phase.
The metamagnetic transition proceeds through nucleation and growth of the
ferromagnetic domains at the expense of the antiferromagnetic phase. The
preferential orientation of the ferromagnetic (tetragonal) domains along the
field direction increases the linear dimension of the sample in the field
direction and decreases in the orthogonal direction leading to the observed
giant anisotropic magnetostriction effect. Our study also suggests that
nanodomains of the low temperature antiferromagnetic phase possibly exist in
the temperature region T$_N$ 
\mbox{$<$}%
T 
\mbox{$<$}%
T$_C$.
\end{abstract}


\section{INTRODUCTION}

Mixed valent RE$_{1-x}^{3+}$A$_x^{2+}$MnO$_3$-type manganites (RE$^{3+}$ and
AE$^{2+}$ are rare earth and alkaline ions respectively) are best known for
their extraordinary sensitivity of resistivity to internal molecular and
external magnetic fields.\cite{Coey} The ferromagnetic ordering of the
localized t$_{2g}^3$ spins at Mn$^{3+}$:t$_{2g}^3$e$_g^1$ and Mn$^{4+}$:t$%
_{2g}^3$e$_g^0$ sites favors the delocalization of e$_g$ carriers due to the
double exchange interaction and, the antiferromagnetic ordering of the t$%
_{2g}^3$ spins favors the localization of the e$_g$-carriers.\cite{Zener}
However, the nature of the antiferromagnetic (AF) ordering in manganites is
strongly influenced by the carrier concentration, the average ionic radius 
\mbox{$<$}%
r$_A$%
\mbox{$>$}%
of the A-site cations and the orbital degree of freedom.\cite{Coey} Hence,
different type of antiferromagnetic configurations (A, pseudo CE, CE, C, and
G) are found in manganites. In particular, the compound Pr$_{0.5}$Sr$_{0.5}$%
MnO$_3$ received much attention due to the discovery of the first order
antiferromagnetic insulator-ferromagnetic metal transition under a magnetic
field.\cite{Tomioka} The low temperature antiferromagnetic transition was
earlier thought to be of CE type, but later studies showed that it is an
A-type antiferromagnet with successive ferromagnetic planes coupled
antiferromagnetically\cite{Kawano}. Charges are itinerant within the
ferromagnetic planes. The A-type antiferromagnetism in this compound is
believed to be the result of the d$_{x^2-y^2}$ orbital ordering of the Mn$%
^{3+}$ ions. It was earlier interpreted that the low temperatue insulating
state in zero field is charge and CE-type antiferromagnetic ordered and, the
destruction of the insulating state under a magnetic field is caused by the
field induced melting of charges.\cite{Tomioka} However, the notion of
charge ordering was later discarded.\cite{Kawano,Damay} Presently, colossal
magnetoresistance in this compound is believed to be the result of the field
induced antiferromagnetic to ferromagnetic transition. However, it is not
known whether the lattice is also affected by the magnetic field and if it
is so in what manner. Studying the nature of the lattice distortion under a
magnetic field could shed light on the mechanism of magnetoresistance. There
are only few reports on the magnetostriction effect in the three dimensional
perovskite manganites\cite{Asamitsu,Ibarra1}. These works mostly concentrate
on the compounds which have ferromagnetic metallic or insulating ground
states. Recently, the magnetostriction behavior of the manganites having
charge and antiferromagnetic (CE-type) orders at low temperature was
investigated.\cite{Mahi1,Mahi2} It was found that the field induced
transition from the charge ordered antiferromagnetic to the charge
delocalized ferromagnetic transition in Nd$_{0.5}$Sr$_{0.5}$MnO$_3$ and La$%
_{0.5}$Ca$_{0.5}$MnO$_3$ \cite{Mahi1,Mahi2} is accompanied by a giant
positive volume magnetostriction (i.e. the lattice volume increases under an
external magnetic field) whereas Nd$_{0.5}$Ca$_{0.5}$MnO$_3$ and some other
charge ordered systems exhibit a giant negative volume magnetostriction
(i.e., lattice volume contracts under an external magnetic field).\cite
{Mahi3} In this context, we investigated the magnetostriction behavior of Pr$%
_{0.5}$Sr$_{0.5}$MnO$_3$ which is neither charge ordered nor exhibits CE
type antiferromagnetism.

\bigskip

\section{EXPERIMENT}

The polycrystalline Pr$_{0.5}$Sr$_{0.5}$MnO$_3$ sample was earlier
characterized by\ X-ray diffraction and zero field resistivity measurements.%
\cite{Rao} The room temperature structure was found to be tetragonal (space
group {\it I4/mcm}) by neutron diffraction.\cite{Woodward} We carried out
magnetization M, Linear thermal expansion ($\Delta $L/L) and
magnetostriction on the previously characterized sample. The temperature
dependence of M at $\mu _0$H = 1mT was measured using a Quantum Design SQUID
magnetometer. We measured the isothermal magnetization up to 12 T using a
vibrating sample magnetometer. The linear thermal expansion $\Delta $L/L by
the strain gauge method \ in absence of \ a magnetic field was measured in
300 K-10 K range. The value of \ $\Delta $L/L at 300 K was taken as the
reference point. Magnetostriction isotherms were measured using a pulsed
magnetic field up to $\mu _0$H = 14 T with the magnetic field parallel and
perpendicular to the measuring direction. From the measured parallel ($%
\lambda _{\parallel }$) and perpendicular ($\lambda _{\perp }$)
magnetostrictions, we calculated the volume ($\omega $) and the anisotropic (%
$\lambda _t$) magnetostrictions for randomly oriented polycrystallites using
the relations $\omega $ = $\lambda _{\parallel }$+2$\lambda _{\perp }$ and $%
\lambda _t$ =$\lambda _{\parallel }$-$\lambda _{\perp }$. The field
dependence of the magnetostriction at selected temperatures was registered
after zero field cooling the sample from 300 K to a desired temperature.

\bigskip

\section{RESULTS}

\bigskip

Figure 1 shows the temperature dependence of M and $\Delta $L/L of Pr$_{0.5}$%
Sr$_{0.5}$MnO$_3$ on the left and right scales respectively. The M(T) curve
suggests that the sample first undergoes a paramagnetic to ferromagnetic
transition (T$_C$ = 270 K) and then from ferromagnetic to antiferromagnetic
with (T$_N$ = 100 K) while cooling. The inflection point of the rapidly
varying part of the M(T) curve was taken as the Neel temperature. The
antiferromagnetic ordering was shown to be A-type.\cite{Kawano} While
warming T$_N$ increases to 125 K and M(T) exhibits hysteresis up to the
Curie temperature. It should be noted that M has not become zero at 10 K (T%
\mbox{$<$}%
\mbox{$<$}%
T$_N$) but exhibits a large value (1/5 of the maximum M(T) while cooling and
1/2 of the maximum M(T) while warming). This suggests that a fraction of the
ferromagnetic phase continues to exist even below T$_N$. The presence of the
ferromagnetic phase below T$_N$ is also shown up in the field dependence of
the magnetization to be shown later. The structural symmetry at room
temperature is tetragonal and it is unaffected by the
paramagnetic-ferromagnetic transition. However, the
ferromagnetic-antiferromagnetic transition is accompanied by tetragonal (%
{\it I4/mcm}, $\sqrt{2}$a$_p$ x$\sqrt{2}$a$_p$x a$_p$) -orthorhombic ({\it %
Fmmm}, 2a$_p$ x 2a$_p$ x 2a$_p$ ,a$_p$ = lattice parameter of a pseudo cubic
unit cell ) structural transition.\cite{Damay} The vertical dashed lines
mark the coexistence of the high and low temperature phases as found by the
neutron diffraction study on a similar composition.\cite{Damay} The $\Delta $%
L/L decreases rather continuously from 300 K down to 10 K without a clear
anomaly either at the Curie temperature or at the Neel temperature. It
should be pointed here that \ a clear volume contraction is observed around
the Neel temperature in Nd$_{0.5}$Sr$_{0.5}$MnO$_3$\cite{Ritter} and La$%
_{0.5}$Ca$_{0.5}$MnO$_{3\text{.}}$\cite{Mahi2} and more recently even in the
higher Sr content compound Pr$_{0.46}$Sr$_{0.54}$MnO$_3$.\cite{Mahi4} The $%
\Delta $L/L of Pr$_{0.5}$Sr$_{0.5}$MnO$_3$ exhibits hysteresis over a wide
temperature range (60 K- 270 K) similar to the M(T) behavior. This behavior
is strikingly different from Pr$_{0.46}$Sr$_{0.54}$MnO$_3$ which showed
hysteresis only around the Neel temperature.\cite{Mahi4}

\smallskip

Figure 2(a) shows the field dependence of the magnetization at selected
temperatures for T 
\mbox{$<$}%
T$_N$ and Figure 2(b) shows the M(H) behavior above T$_N$. The behavior the
M(H) curve below T$_N$ suggests the coexistence of the ferromagnetic and the
antiferromagnetic phases. The behavior of M(H) below $\mu _0$H = 2 T at 25 K
is dominated by the domain wall displacement and domains alignment of the
ferromagnetic phase. The sudden increase of M(H) just above $\mu _0$H$_C$ =
4 T is the result of the metamagnetic transition in the antiferromagnetic
phase. A rough value of the ferromagnetic phase fraction can be estimated as 
$y$ = $M_0$/M$_{FM}$ where M$_0$ is the spontaneous magnetization and M$%
_{FM} $ is the saturation magnetization if the whole sample is
ferromagnetic. The value of M$_0$ can be determined by plotting M versus 1/H
and extrapolating the linear part of the high field magnetization ($\mu _0$H 
\mbox{$<$}%
2 T) to 0 T which we found to be 0.46 $\mu _B$. The value of M$_{FM}$ for
this composition is 3.5 $\mu _B$ and hence y $\approx $ 13 \%. The volume
fraction of the ferromagnetic phase increases when the sample undergoes the
metamagnetic transition and finally y reaches 100 \% at $\mu _0$H = 12 T. As
T increases from 25 K, $\mu _0$H$_C$ decreases from 4 T to 1.05 T at 100 K.
The size of the ferromagnetic phase in the antiferromagnetic host appears to
be dependent on the sample preparation condition. Allodi et al.\cite{Allodi}
and Jung et al.\cite{Jung} suggested nanoscale size ferromagnetic clusters
at 10 K ( 
\mbox{$<$}%
\mbox{$<$}%
T$_N$) whereas our results suggest rather a macroscopic (few 1000 Angstrom)
size ferromagnetic phase. Very recently, Boujelben et al.\cite{Boujelben}
showed that the magnetic properties of Pr$_{0.5}$Sr$_{0.5}$MnO$_3$ is
sensitive to the quenching condition. The sample air quenched to room
temperature from 1673 K was orthorhombic ({\it Imma}) and underwent
ferromagnetic-antiferromagnetic transition whereas the water quenched sample
was rhombohedral ({\it R3c})) and did not show antiferromagnetic transition.
Both the samples were found to be stoichiometric and the differences in
magnetic properties were suggested to disorder effect in A-site.\cite
{Boujelben} Our sample was furnace cooled (5 K/min) from 1573 K to room
temperature and was found to be tetragonal.\cite{Rao,Woodward} Recent
theoretical model predict that the strength of disorder at A-site cations
can control the size of two coexisting phases.\cite{Dagotto}

\bigskip

\smallskip Fig. 3(a) shows the parallel magnetostriction ($\lambda
_{\parallel }$) isotherms for T $\leq $ T$_N$ and Fig. 3(c) shows $\lambda
_{\parallel }$ for T$_N$ 
\mbox{$<$}%
T 
\mbox{$<$}%
T$_C$. \ At 25 K, $\lambda _{\parallel }$ is small below $\mu _0$H$_C$ = 4 T
and then increases rapidly in between 4 T and 6 T. A change of slope occurs
around 6.5 T and above this field $\lambda _{\parallel }$ increases less
rapidly without saturation up to the maximum field of 14.2 T. The value of $%
\lambda _{\parallel }$ at 14.2 T is 900 x 10$^{-6}$. A clear hysteresis is
seen while reducing the field. When the field is reduced to 0 T, $\lambda
_{\parallel }$ does not return to the original starting value but \ attains
a higher value ($\approx $50 x 10$^{-6}$). The rapid increase of $\lambda
_{\parallel }$ just above 4 T correlates with the metamagnetic transition
found in the M(H) behavior (see Fig. 2). The metamagnetic like behavior in
magnetostriction is present until 100 K and as T increases from 25 K the
metamagnetic like transition occurs at lower fields which is in agreement
with M(H) behavior (Fig. 2). At 125 K, 150 K and 175 K, $\lambda _{\parallel
}$ increases rapidly at very low fields ( $\mu _0$H$_C$ 
\mbox{$<$}%
0.5 T) due to ferromagnetic domain wall motion. However, $\lambda
_{\parallel }$ increases without saturation up to the maximum field. We do
not see any ferromagnetic contribution to the magnetostriction below the
Neel temperature although such a behavior is shown in M(H) (see Fig. 2).
This is possibly due to the smaller fraction of the ferromagnetic phase.
Fig. 3(b) and Fig. 3(d) shows the perpendicular magnetostriction ($\lambda
_{\perp }$) isotherms. As for $\lambda _{\parallel }$, $\lambda _{\perp }$
undergoes a rapid change during the metamagnetic transition but the sign is
opposite to $\lambda _{\perp }$. An important difference is that the
magnitude of the maximum magnetostriction value is lower in the
perpendicular case ($\lambda _{\perp }$ = 600 x 10$^{-6}$, $\lambda
_{\parallel }$ = 900 x 10$^{-6}$ at 14.2 T and T = 25 K) and the hysteresis,
in particular the irreversibility at the origin is much larger in the
perpendicular magnetostriction isotherms. For example, $\lambda _{\perp }$ =
-200 x 10$^{-6}$ and $\lambda _{\parallel }$ = 50 x 10$^{-6}$ at 0 T after
the field is reduced from the highest value. The irreversibility vanishes
above the Neel temperature (see 125 K data in Fig. 2(b) and Fig. 2(d)).

\bigskip

Figs. 4(a) and 4(b) show the anisotropic magnetostriction ($\lambda _t$)
isotherms calculated from the parallel and perpendicular magnetostrictions
in Figs. 3 (a)-(d). The anisotropic magnetostriction is nearly zero in the
antiferromagnetic state but increases rapidly during the metamagnetic
transition. At T = 25 K and $\mu _0$H = 14.2 the anisotropic
magnetostriction reaches the maximum value of 1500 x 10$^{-6}$. It is indeed
surprising since such a huge value anisotropic magnetostriction is not found
in manganites so far. In a number of manganites\cite{Ibarra1}, the
anisotropic magnetostriction was found to be negligible ($\lambda _t$ 
\mbox{$<$}%
60 x 10$^{-6}$) compared to the volume magnetostriction ($\approx $ 10$^{-2}$%
). What is also surprising is that the anisotropic magnetostriction even in
between T$_N$ and T$_C$ (see 125 K, 150 K, and 175 K data) is at least a
factor of 3-5 times larger than what is found in the ferromagnetic manganite
La$_{0.7}$Ca$_{0.3}$MnO$_3$ in the same temperature range.\cite{Teresa1} The
field dependence of $\lambda _t$ below 125 K shows a strong irreversibility
at the origin which is a reflection of the behavior of the perpendicular and
the parallel magnetostriction isotherms in Fig. 3. In Pr$_{0.46}$Sr$_{0.54}$%
MnO$_3$, the maximum $\lambda _t$ was less than 250 x 10$^{-6}$ at 14.2 T.%
\cite{Mahi4} The volume magnetostriction ($\omega $) shows a complex
behavior as a function of field. The value of $\omega $ at T = 25 K is
negligible in the antiferromagnetic state and shows a sudden decrease during
the metamagnetic transition. At T = 25 K, $\omega $ reaches -250 x10$^{-6}$
at the maximum field which is about six times lower than the $\lambda _t$
value at the corresponding field. When the field is decreased from the
maximum value $\omega $ decreases to reach still a lower value -600 x 10$%
^{-6}$ at 1.5 T and then increases again for further decrease in H. At the
origin, $\omega $ is much lower (-500 x 10$^6$) with respect to the starting
value. There is not much a big change in $\omega $ value between 25 K and
100 K but $\omega $ becomes positive above 125 K as can be clearly seen in
Fig. 4(c) and Fig. 4(d). It is noteworthy to mention that the volume
magnetostriction is positive and larger in value ($\omega $ 
\mbox{$>$}%
1000 x 10$^{-6}$ at 25 K) below the Neel temperature in other half doped
manganites Nd$_{0.5}$Sr$_{0.5}$MnO$_3$ and La$_{0.5}$Ca$_{0.5}$MnO$_3$.\cite
{Mahi1,Mahi2} and also inThe Pr$_{0.46}$Sr$_{0.54}$MnO$_3$.\cite{Mahi4}

\bigskip

Fig. 5 shows the value of $\lambda _t$ and $\omega $ at the maximum field ($%
\mu _0$H = 14.2 T) as a function of the temperature. A rapid change in both $%
\lambda _t$ and $\omega $ occurs as the Neel temperature is approached from
below. The anisotropic magnetostriction decreases rapidly and the volume
magnetostriction changes the sign as T$_N$ is approached. No clear anomaly
is observed at the Curie temperature in contrast to the behavior of La$%
_{0.7} $Ca$_{0.3}$MnO$_3$.\cite{Teresa1} The lack of volume anomaly at T$_C$
can be understood as a result of weak electron-phonon interaction in the
paramagnetic state compared to La$_{0.7}$Ca$_{0.3}$MnO$_3$. It is worthy to
note that charges are itinerant in the paramagnetic state (d$\rho $/dT 
\mbox{$>$}%
0) of our compound whereas they are localized in La$_{0.7}$Ca$_{0.3}$MnO$_3$
(d$\rho $/dT 
\mbox{$<$}%
0).

\section{\protect\bigskip DISCUSSIONS}

What is the origin of the unusual giant anisotropic magnetostriction in Pr$%
_{0.5}$Sr$_{0.5}$MnO$_3$ ? Giant anisotropic magnetostriction is generally
found in rare earth intermetallic compounds like TbFe$_2$ and it is caused
by the coupling of magnetic moments to the oval or pancake shaped
anisotropic charge cloud of \ 4f orbitals of the rare earth ion.\cite{Clark}
The anisotropic magnetostriction in our compounds due to spin-orbit coupling
of Pr-ions can be neglected since anisotropic effect was not observed in
related compounds.\cite{Teresa2} We have to consider other possibilities.
The A-type antiferromagnetism in Pr$_{0.5}$Sr$_{0.5}$MnO$_3$ is caused by
the e$_g$-d$_{x^2-y^2}$orbital ordering of Mn$^{3+}$ ions which drives the
tetragonal ({\it I4/mcm}) to orthorhombic ({\it Fmmm}) structural
transition. The e$_g$-orbital moment is quenched but orbital ordering
creates a quadrupole moment.\cite{Kugel} Since there is an unpaired electron
in the d$_{x^2-y^2}$ orbital, there exists a magnetic dipole moment in
addition to the quadrupole moment. The quadrupole moment (which is also
assigned to pseudo spin $\tau ^z$ = 1/2 for d$_{3z^2-r^2}$ and $\tau ^z$ =
-1/2 for d$_{x^2-y^2}$ orbitals)\cite{Kugel} of the e$_g$-orbital can
interact with the lattice through the quadrupole-strain interaction. When
the spin ordering changes from the A-type antiferromagnetic to
ferromagnetic, the d$_{x^2-y^2}$ orbital ordering is also expected to be
destroyed. Hence, a magnetic field induced structural transition from {\it %
Fmmm} to {\it I4/mcm} symmetry takes place. The field induced transition is
of first order with coexistence of the A-type AF domains ({\it Fmmm}) and
the ferromagnetic domains ({\it I4/mcm}). At very high field (H 
\mbox{$>$}%
\mbox{$>$}%
H$_C$), only the ferromagnetic mono domain(I4/mcm) is expected.\ However,
the orientation and perhaps the shape of the crystallographic ({\it I4/mcm})
domains just above H$_C$ depend on the direction of the applied field. When
the measuring direction is parallel to the applied magnetic field, the {\it %
I4/mcm} domains are favorably oriented in the field direction and hence the
parallel magnetostriction is positive and its value is larger. If we
consider columns of the {\it I4/mcm} domains aligned in the field direction,
the sign of the perpendicular magnetostriction indicates that the columns
contract in the lateral direction. Hence, the anisotropic magnetostriction
which is the difference between the parallel and perpendicular
magnetostrictions is large. When the field is reduced from the high value,
some of the {\it I4/mcm} domains do not revert back to the {\it Fmmm}
structure which causes the irreversibility at the origin. The rapid decrease
of the anisotropic magnetostriction as T$_N$ is approached from below can be
understood as a result of the decrease in the phase fraction of the {\it Fmmm%
} phase.

\smallskip \smallskip

The large value of the anisotropic magnetostriction (see 125 K, 150 K and
175 K data) above the Neel temperature but below the Curie temperature is
intriguing. A possible scenario is that nano domains of the low temperature
orbital ordered antiferromagnetic phase already exist in the high
temperature ferromagnetic phase. As the temperature reduces these
nanodomains grow in size and fuse into macrodomains around the Neel
temperature. This can cause hysteresis in M(T) and $\Delta $L/L as we have
observed. Such a scenario is not an unlikely possibility given the fact that
we also observed appreciable positive volume magnetostriction in the
ferromagnetic phase of Nd$_{0.5}$Sr$_{0.5}$MnO$_3$ and we have suggested the
possibility of the charge-orbital ordered domains in the ferromagnetic phase 
\cite{Mahi1}. More recently, orbital ordered nanostructure was indeed found
by X-ray diffusive scattering study in the same composition.\cite{Kiryukhin}
Then, the contribution from the field induced structural changes of these
nanodomains adds to the usual ferromagnetic contribution and hence the
anisotropic magnetostriction is large. Finally, some remarks to be made
about the differences between the magnetostriction behaviors of Pr$_{0.46}$Sr%
$_{0.54}$MnO$_3$ and Pr$_{0.5}$Sr$_{0.5}$MnO$_3$. It was shown earlier\cite
{Mahi4} that Pr$_{0.46}$Sr$_{0.54}$MnO$_3$ exhibits a giant positive volume
magnetostriction in constrast to the small negative volume magnetostriction
effect found in Pr$_{0.5}$Sr$_{0.5}$MnO$_3$. But, the anisotropic
magnetostriction in Pr$_{0.46}$Sr$_{0.54}$MnO$_3$ is an order of magnitude
smaller than in Pr$_{0.5}$Sr$_{0.5}$MnO$_3$. We believe that these
differences are related to the volume of differences between the
orthorhombic(antiferromagnetic) and tetragonal (ferromagnetic or
paramagnetic) phases. The spontaneous volume ( H = 0 T) shows a dramatic
decrease around the Neel temperature in Pr$_{0.46}$Sr$_{0.54}$MnO$_3$
whereas it is rather smooth across the ferromagnetic-antiferromagnetic
transition in Pr$_{0.5}$Sr$_{0.5}$MnO$_3$. The behavior of the d$_{x^2-y^2}$
orbital ordering due to excess electrons and the magnitude of exchange
interactions will play important role in these two compounds, but the
details are yet to be understood.

\smallskip

\section{SUMMARY}

To summarize, the linear thermal expansion of Pr$_{0.5}$Sr$_{0.5}$MnO$_3$
does not show any clear anomaly either at the ferromagnetic or
antiferromagnetic transition. Our isothermal magnetization study suggests\
that a minority ferromagnetic phase coexists with the majority
antiferromagnetic phase below the Neel temperature. The ferromagnetic phase
fraction is about 13 \% at 25 K. The field induced antiferromagnetic to
ferromagnetic transition is accompanied by a giant anisotropic
magnetostriction ($\lambda _t$ $\approx $ 10$^{-3}$) and a smaller volume
contraction ($\omega $ $\approx $10$^{-4}$). This is the first time such a
large anisotropic magnetostriction is found in manganites. The value of
anisotropic magnetostriction is large below the Neel temperature but its
value even in the ferromagnetic temperature region is larger than found in
other manganites. The anisotropic magnetostriction does not saturate even at
14.2 T in the ferromagnetic temperature region. We suggest that the
metamagnetic transition in spin sector is accompanied by the creation of
tetragonal ($I4/mcm$) ferromagnetic domains from the orthorhombic ($Fmmm$)
antiferromagnetic matrix. The large anisotropic magnetostriction is
suggested to formation of\ ferromagnetic domains ($I4/mcm$) in the field
direction. Our magnetostriction and temperature dependence of the
magnetization also suggest that nanodomains of the low temperature $Fmmm$
(antiferromagnetic phase) are possibly present in 125 K- 200 K where the
sample is supposed to be a long range ferromagnet. Contrary to the large
anisotropic magnetostriction, the volume magnetostriction is rather small.
In view of the large anisotropic magnetostriction found it is worth to
investigate anisotropic properties in magnetoresistance and magnetization as
was found in the case of La$_{1-x}$Sr$_x$CoO$_3$.\cite{Mahi5}

\smallskip

\section{ACKNOWLEDGMENTS}

\bigskip

R. M thanks MENRT (France) and Ministrio de Ciencia y Cultura (Spain) for
financial assistance.

\smallskip

* Present Address : Department of Physics, 104 Davey Laboratory, Box 54, The
Pennsylvania State University, University Park, PA-16802-6300, USA

e-mail: mur5@psu.edu

\begin{center}
\bigskip \newpage FIGURE\ CAPTIONS
\end{center}

\begin{description}
\item[Fig. 1]  : Temperature dependence of magnetization (left scale) and \
zero- field linear thermal expansion (right scale) of Pr$_{0.5}$Sr$_{0.5}$MnO%
$_3$ during cooling from 300 K to 10 K \ warming back to 300 K. M(T) was
measured under $\mu _0$H = 1 $mT$.

\item[Fig. 2]  : Field dependence of magnetization (a) below T$_N$ and (b)
above T$_N$ in Pr$_{0.5}$Sr$_{0.5}$MnO$_3$

\item[Fig. 3]  : Parallel and Perpendicular magnetostriction isotherms of Pr$%
_{0.5}$Sr$_{0.5}$MnO$_3$

\item[Fig. 4]  Anisotropic and Volume magnetostriction isotherms of Pr$_{0.5}
$Sr$_{0.5}$MnO$_3$.

\item[Fig. 5]  : Temperature dependence of volume ($\omega $) and
anisotropic ($\lambda _t$) magnetostriction at $\mu _0$H = 14.2 T for Pr$%
_{0.5}$Sr$_{0.5}$MnO$_3$.
\end{description}

\end{document}